\documentclass{elsart3p}

\usepackage{epsfig}

\usepackage{amssymb}
\usepackage{multicol}

\newcommand{\ee}{{\mathrm{e}}}
\newcommand{\dotn}{{\dot{n}}}

\begin{document}

\begin{frontmatter}



\title{Comment on ``Delayed luminescence of biological systems
  in terms of coherent states''
  [Phys. Lett. A 293 (2002) 93]}

\author{Vahid Salari},
\ead{vahid.salari@impmc.upmc.fr} 
\author{Christian Brouder},
\address{Institut de Min\'eralogie et de Physique des 
  Milieux Condens\'es,
Universit\'e Pierre et Marie Curie-Paris 6,
CNRS UMR7590,
Universit\'e Denis Diderot-Paris 7,
Institut de Physique du Globe,\\
Campus Jussieu, bo{\^{\i}}te courrier 115,
4 place Jussieu, 75252 Paris cedex 05, France.
}%

\begin{abstract}
Popp and Yan [F. A. Popp, Y. Yan, Phys. Lett. A 293 (2002) 93]
proposed a model for delayed luminescence
based on a single time-dependent coherent state.
We show that 
the general solution of their model corresponds to
a luminescence that is a linear function of time.
Therefore, their model is not compatible with any
measured delayed luminescence.
Moreover, the functions that they use to describe
the oscillatory behaviour of delayed luminescence
are not solutions of the coupling equations 
to be solved.
\end{abstract}

\begin{keyword}
Delayed luminescence; Hyperbolic oscillations; Coherent states;
Biophotonics
\PACS 42.50.-p\sep 78.55.-m
\end{keyword}
\end{frontmatter}

\section{Introduction}
Popp and Yan proposed a model for delayed luminescence
based on a single time-dependent coherent state~\cite{Popp-02}.
The standard explanation of delayed luminescence
from plants is made in terms of 
Photosystem~II reaction centers~\cite{Goltsev-09,Guo-09},
but the Popp and Yan approach is often considered to be
a possible alternative
(see Refs.\cite{Katsumata-08,Lanzano-09,Winkler-09,Kun-09}
for recent references).

Moreover, the assumed validity of the Popp and Yan
model is used as a confirmation of
a very speculative theory claiming
that light is used by cells for organizational
tasks (''[\dots] the
capacity of living systems to trap light and
to use it for organizational tasks''~\cite{Yan-05}).

These applications make it useful to check
the validity of the model of Popp and Yan.
In this comment, we point out 
that the hyperbolic decay of delayed luminescence
cannot be described by the Popp and Yan model, 
and that the function they obtain to describe
the oscillations of delayed luminescence is not
consistent with their coupling equations.

\section{Hyperbolic relaxation}
The primed equations refer to ref.~\cite{Popp-02}.
The time-dependent Hamiltonian given by
eq.~(2') has the solution $|\alpha\rangle$,
which is a coherent state, i.e. an
eigenstate of the annihilation operator
$a(t)$ (in the Heisenberg picture) such as
$a(t)|\alpha\rangle=\alpha(t)|\alpha\rangle$,
where $\alpha(t)$ is a complex function of $t$.
The general form of $\alpha(t)$ is
(correcting two misprints in eq.~(5')),
\begin{eqnarray}
\alpha(t) &=& \ee^{-i\psi(t)} 
\Big(\alpha(0) 
  -i \int_0^t f(t')\ee^{i\psi(t')} d t'\Big).
\label{alphat}
\end{eqnarray}

The number of photons as a function
of time is given by
$n(t)=\langle \alpha| a^+(t) a(t)|\alpha\rangle
=|\alpha(t)|^2$. Thus,
\begin{eqnarray}
n(t) &=& 
\Big|\alpha(0) 
  -i \int_0^t f(t')\ee^{i\psi(t')} d t'\Big|^2.
\label{nt}
\end{eqnarray}

By assuming ``homeostasis'' and 
the fact that ``$\alpha$ should not
be influenced by external classical energy
sources'' (we do not comment here on the validity
of these assumptions), Popp and Yan
obtain eq.~(9') for $f$, with general
solution 
$f(t)=f(0) \ee^{-i \psi(t)}$.
Note that, at this stage, Popp and Yan consider
a special $\omega(t)$ while we use a general one.
By introducing this value of $f(t)$ into
eq.~(\ref{nt}), we obtain
\begin{eqnarray}
n(t) &=& 
|\alpha(0) 
  -i f(0) t|^2.
\label{ntlin}
\end{eqnarray}
Therefore, $n(t)$ is a quadratic function
of $t$, independent of $\omega(t)$.

The relation between the intensity of
light $I(t)$ and $n(t)$ is not given
explicitly in Ref.~\cite{Popp-02}, but the caption
of Fig.~3 shows that the authors
use the relation $I(t)\propto \dotn(t)$.
It follows from eq.~(\ref{ntlin}) that $I(t)$ is a linear
function of time. This does not agree
with any measurement of delayed luminescence.

Therefore, the experimental hyperbolic relaxation given in eq.~(17') is
not a consequence of the coherent-state model of the paper.

\section{Oscillations}

Popp and Yan claim that the oscillatory behavior of
delayed luminescence can be explained by a coupling
of two coherent states $|\alpha_1\rangle$ and
$|\alpha_2\rangle$ described by differential equation (18').
We shall see that this interpretation meets 
a rather serious inconsistency.
In their calculation, they use eq.~(18') to derive eq.~(21')
and solve eq.~(21') with $\alpha_1$ and $\alpha_2$
defined by eq.~(22'). However, these
$\alpha_1$ and $\alpha_2$ are not solutions of the
starting eq.~(18'). In other words, their ``solutions''
do not solve the coupling equation they assume.
The point is that a solution of eq.~(21')
is generally not a solution of eq.~(18'). 
Indeed, take any differentiable real function $y_1(t)$
and define $y_2=y+y_1$, with 
\begin{eqnarray*}
y(t) &=& \kappa\mathrm{ln}(1+\lambda_1 t)-
\kappa\mathrm{ln}(1+\lambda_2 t)+\phi.
\end{eqnarray*}
Then, $\alpha_1(t)=|a_1| \ee^{-i y_1(t)}$ and
$\alpha_2(t)=|a_2| \ee^{-i y_2(t)}$ define a solution
of eq.~(21') which is not a solution of eq.~(18').

\section{Conclusion}
The main purpose of this comment was to show that the
coherent state model proposed by Popp and Yan 
does not agree with delayed luminescence experiments:
the math is simply wrong.

Other points of the paper could also have been discussed.
For instance, the density corresponding to
a linear combination of coherent states
is not given by $n(t)=|\alpha_1(t)+\alpha_2(t)|^2$
(see ref.~\cite{MandelWolf}),
the question why the two coherent states should have the
same weight, and so on. But we think that our argument
is strong enough already.


\begin{thebibliography}{1}

\bibitem{Popp-02}
F.~A. Popp and Y.~Yan.
\newblock Delayed luminescence of biological systems in terms of coherent
  states.
\newblock {\em Phys. Lett. A}, 293:93--7, 2002.

\bibitem{Goltsev-09}
V.~Goltsev, I.~Zaharieva, P.~Chernev, and R.~J. Strasser.
\newblock Delayed luminescence in photosynthesis.
\newblock {\em Photosynthesis research}, 101:217--32, 2009.

\bibitem{Guo-09}
Y.~Guo and J.-L. Tan.
\newblock A kinetic model structure for delayed fluorescence from plants.
\newblock {\em BioSystems}, 95:98--103, 2009.

\bibitem{Katsumata-08}
M.~Katsumata, A.~Takeuchi, K.~Kazumura, and T.~Koike.
\newblock {New feature of delayed luminescence: preillumination-induced
  concavity and convexity in delayed luminescence decay curve in the green alga
  {\it{Pseudokirchneriella subcapitata}}.}
\newblock {\em J. Photochem. Photobiol. B}, 90:152--62, 2008.

\bibitem{Lanzano-09}
L.~Lanzan\`{o}, Li~Sui, E.~Costanzo, M.~Gulino, A.~Scordino, S.~Tudisco, and
  F.~Musumeci.
\newblock {Time-resolved spectral measurements of delayed luminescence from a
  single soybean seed: effects of thermal damage and correlation with
  germination performance.}
\newblock {\em Luminescence}, 24:409--15, 2009.

\bibitem{Winkler-09}
R~Winkler, H~Guttenberger, and H~Klima.
\newblock {Ultraweak and induced photon emission after wounding of plants.}
\newblock {\em Photochemistry and photobiology}, 85(4):962--5, 2009.

\bibitem{Kun-09}
Si~Kun, Chun-Li Liu, and Jia Hun-Yu.
\newblock Biophoton emission of biological systems in terms of odd and even
  coherent states.
\newblock In Xun Shen, Xiao-Lin Yang, Xin-Rong Zhang, Zhong-Jie Cui, L.~J.
  Kricka, and P.~E. Stanley, editors, {\em Proceedings of the 15th 
  International
  Symposium on Bioluminescence and Chemiluminescence}, pages 63--6, Singapore,
  2009. World Scientific.

\bibitem{Yan-05}
Yu~Yan, F.-A. Popp, S.~Sigrist, D.~Schlesinger, A.~Dolf, Zhongchen Yan,
  S.~Cohen, and A.~Chotia.
\newblock {Further analysis of delayed luminescence of plants.}
\newblock {\em J. Photochem. Photobiol. B}, 78:235--44, 2005.

\bibitem{MandelWolf}
L.~Mandel and E.~Wolf.
\newblock {\em Optical Coherence and Quantum Optics}.
\newblock Cambridge University Press, Cambridge, 1995.

\end{thebibliography}
\end{document}